\documentclass[a4paper,12pt]{article}
\usepackage{epsfig,amsmath}

\newcommand{\bs}{\boldsymbol}
\newcommand{\pd}{\partial}

\begin{document}

\raggedbottom

\title{Semiclassical spin coherent state method \\ in the weak spin-orbit
coupling limit}
\author{Oleg Zaitsev}
\date{{\em Institut f\"ur Theoretische Physik, Universit\"at Regensburg,
D-93040 Regensburg, Germany}}
\maketitle

\abstract{We apply the semiclassical spin coherent state method for the density
of states by Pletyukhov {\em et al.} (2002) in the weak spin-orbit coupling
limit and recover the modulation factor in the semiclassical trace formula
found by Bolte and Keppeler (1998, 1999).}
\vspace*{1cm}

\section{Introduction}

A new solution to the problem of how to include spin-orbit interaction in the
semiclassical theory was recently proposed by Pletyukhov {\em et al.}
\cite{Ple}. They use the spin coherent states to describe the spin degrees of
freedom of the system. Then a path integral that combines the spin and orbital
variables can be constructed, leading to the semiclassical propagator (or its
trace) when evaluated within the stationary phase approximation. In such an
approach the spin and orbital degrees of freedom are treated on equal footings.
In particular, one can think of a classical trajectory of the system in the
extended phase space, i.e., the phase space with two extra dimensions due to
spin. (The spin part of the extended phase space can be mapped onto the Bloch
sphere.) Like in the pure orbital systems, it is possible to construct a
classical Hamiltonian that will be a function of the phase space coordinates.
The trajectories of the system satisfy the equations of motion generated by
this Hamiltonian.

In this letter we apply the general theory \cite{Ple} to the limiting case of
weak spin-orbit coupling. This limit is naturally incorporated in the theory
proposed by Bolte and Keppeler \cite{Bol} that based on the $\hbar \rightarrow
0$ expansion in the Dirac (or Schr\"odinger) equation. Bolte and Keppeler have
shown that the semiclassical trace formula without spin-orbit interaction
acquires an additional modulation factor due to spin, but otherwise remains
unchanged. We obtain the same modulation factor using the spin coherent state
method.

\section{Classical dynamics and periodic \\ orbits} \label{ClasDyn}

We begin with the classical phase space symbol of the Hamiltonian \cite{Ple}
\begin{equation}
H(p,q,z) = H_0 (p,q) + \kappa \hbar S \bs{\sigma}(z) \cdot \mathbf{C}(p,q)
\equiv H_0 + \hbar H_{\mathrm{so}}.
\label{Ham}
\end{equation}
It includes the spin-orbit interaction term $\hbar H_{\mathrm{so}}$ which is
linear in spin, but otherwise is an arbitrary function of (possibly
multidimensional) momenta and coordinates $p$ and $q$. The spin direction is
described by a unit vector $\bs \sigma (z) \overset {\mathrm {def}} = \left< z
\right| \bs{\hat \sigma} \left| z \right>$, where $\bs{\hat \sigma}$ are the
Pauli matrices and the complex variable $z \equiv u - iv$ labels the spin
coherent states of total spin $S$ \cite{Koc}. At the end of our calculations we
will set $S= \frac 1 2$. The Planck constant appears explicitly in this
classical Hamiltonian and is treated as the perturbation parameter in the
weak-coupling limit. The spin-orbit coupling strength $\kappa$ is kept finite.
Thus the condition $\hbar \rightarrow 0$ provides both the semiclassical (high
energy) and the weak-coupling limits. 

The Hamiltonian (\ref{Ham}) determines the classical equations of motion for
the orbital and spin degrees of freedom \cite{Ple}
\begin{eqnarray}
\dot p &=& - \frac {\pd H} {\pd q} = - \frac {\pd H_0} {\pd q} - \kappa \hbar S
\bs{\sigma} \cdot \frac {\pd \mathbf{C}} {\pd q},
\label{EOM1}\\
\dot q &=& \frac {\pd H} {\pd p} = \frac {\pd H_0} {\pd p} + \kappa \hbar S
\bs{\sigma} \cdot \frac {\pd \mathbf{C}} {\pd p},
\label{EOM2}\\
\dot {\bs{\sigma}} &=& - \kappa \bs{\sigma} \times \mathbf{C}.
\label{EOM3}
\end{eqnarray}
Since 
\begin{equation}
\bs{\sigma} (z) = \frac 1 {1 + |z|^2} \left(2u, 2v, |z|^2 -1\right)^{\mathrm T}
\label{sigma}
\end{equation}
in the ``south'' gauge,\footnote{By the south gauge we mean the choice of
parameterization of the spin coherent states by $z$ such that $\sigma_z (|z|
\rightarrow \infty) = 1 $.} Eq.\ (\ref{EOM3}) is equivalent to two
Hamilton-like equations 
\begin{align}
&\dot u = - \frac {\left(1 + |z|^2\right)^2} {4 \hbar S} \frac {\pd H} {\pd v}
= - \frac {\kappa} 4 \left(1 + |z|^2\right)^2 \frac {\pd \bs{\sigma}} {\pd v}
\cdot \mathbf{C},
\label{EOMu} \\
&\dot v = \frac {\left(1 + |z|^2\right)^2} {4 \hbar S} \frac {\pd H} {\pd u} =
\frac {\kappa} 4 \left(1 + |z|^2\right)^2 \frac {\pd \bs{\sigma}} {\pd u} \cdot
\mathbf{C}.
\label{EOMv}
\end{align}
In the leading order in $\hbar$ we keep only the unperturbed terms in Eqs.\ 
(\ref{EOM1}) and (\ref{EOM2}). It follows then that the orbital motion, in this
approximation, is unaffected by spin. The spin motion is determined by the
unperturbed orbital motion via Eq.\ (\ref{EOM3}), which does not contain
$\hbar$. It describes the spin precession in the time-dependent effective 
magnetic field $\mathbf{C} \left(p_0(t), q_0 (t) \right)$, where $\left(p_0(t),
q_0 (t) \right)$ is an orbit of the unperturbed Hamiltonian $H_0$.

\begin{figure}[tbh]
\begin{center}
{\hspace*{0cm}
\psfig{figure=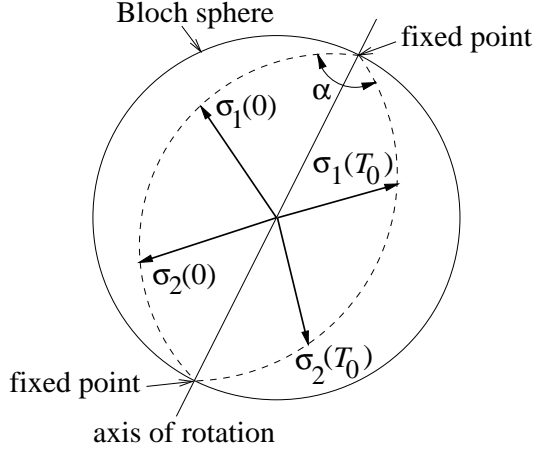,height=5.9cm,width=7cm,angle=0}}
\end{center}
{\vspace*{0cm}}
\caption{Axis of rotation and fixed points of the map $\bs{\sigma} (0)
\longmapsto \bs{\sigma} (T_0)$.} 
\label{Fig1}
{\vspace{.5 cm}}
\end{figure}

In order to apply a trace formula for the density of states, we need to know
the periodic orbits of the system, both in orbital and spin phase space
coordinates. The orbital part of a periodic trajectory is necessarily a
periodic orbit of $H_0$. Assume that such an orbit with period $T_0$ is given.
Then Eq.\ (\ref{EOM3}) generates a map on the Bloch sphere $\bs{\sigma} (0)
\longmapsto \bs{\sigma} (T_0)$ between the initial and final points of a spin
trajectory $\bs{\sigma} (t)$. The fixed points of this map correspond to
periodic orbits with the period $T_0$. Since Eq.\ (\ref{EOM3}) is linear in
$\bs{\sigma}$, for any two trajectories $\bs{\sigma_1} (t)$ and $\bs{\sigma_2}
(t)$, their difference also satisfies this equation. But this means that
$|\bs{\sigma_1} (t) - \bs{\sigma_2} (t)| = \mathrm{const}$, i.e., the angles
between the vectors do not change during the motion. Hence the map is a
rotation by an angle $\alpha$ about some axis through the center of the Bloch
sphere. The points of intersection of this axis and the sphere are the fixed
points of the map (Fig.\ \ref{Fig1}). Thus for a given periodic orbit of $H_0$,
there are two periodic orbits of $H$ with opposite spin orientations (unless
$\alpha$ is a multiple of $2\pi$, by accident). The angle $\alpha$ was
mentioned in Ref.\ \cite{Kep}.
\vspace*{.5cm}

\section{Modulation factor}

\subsection{Correction to the action}

In order to derive a modulation factor in the trace formula, we need to
determine the correction to the action and the stability determinant due to the
spin-orbit interaction. The action along a periodic orbit is~\cite{Ple}
\begin{equation}
{\cal S} = \oint p dq + 2S \hbar \oint \frac {u dv - v du} {1 + |z|^2} 
\equiv {\cal S}_{pq} + \hbar {\cal S}_{\mathrm{spin}}.
\end{equation}
While the spin part contains $\hbar$ explicitly, we need to extract the leading
order correction to the orbital action. This is the only place where we
implicitly take into account the influence of spin on the orbital motion. It is
convenient for the following calculation to parameterize both the perturbed and
unperturbed orbits by a variable $s \in [0,1]$, i.e.,
\begin{equation}
{\cal S}_{pq} = \int_0^1 p \frac {dq} {ds} ds.
\end{equation}
The time parameterization would be problematic since the periods of the
perturbed and the unperturbed orbits differ by order of $\hbar$ (see Appendix
\ref{time}). The correction to the orbital part due to the perturbation~is
\begin{eqnarray}
\delta {\cal S}_{pq} &=& \int_0^1 \left[ \delta p \frac {dq_0} {ds} + p_0 \frac d
{ds} (\delta q) \right] ds
\nonumber \\ 
&=& \int_0^1 \left( \delta p \frac {dq_0} {ds} - \delta q \frac {dp_0} {ds}
\right) ds +  p_0 \delta q \bigg|_0^1.
\label{deltaSpq}
\end{eqnarray}
The boundary term vanishes for the periodic orbit, and the integration can be
done over the period of the unperturbed orbit now:
\begin{eqnarray}
\delta {\cal S}_{pq} &=& \int_0^{T_0} \left( \delta p \dot q_0 - \delta q \dot
p_0 \right) dt 
\nonumber \\ 
&=& \int_0^{T_0} \left( \delta p \frac {\pd H_0} {\pd p} + \delta
q \frac {\pd H_0} {\pd q} \right) dt = \int_0^{T_0} \delta H_0 dt.
\end{eqnarray}
Since the perturbed and unperturbed orbits have the same energy, the variation
of the Hamiltonian is $\delta H_0 = - \hbar H_{\mathrm{so}}$. Taking into
account Eq.\ (\ref{sigma}), we can express the change in the orbital action as
\begin{equation}
\delta {\cal S}_{pq} = - \hbar \kappa S \int_0^{T_0} \mathbf{C} \cdot \bs
\sigma dt = - \hbar \kappa S \int_0^{T_0} \mathbf{C} \cdot \left(
\begin{array}{c}
2u \\ 2v \\ |z|^2 -1
\end{array}
\right) \frac {dt} {1+|z|^2}.
\label{orb}
\end{equation}

We now turn to the spin action. Parameterizing the trajectory with time and
then using the equations of motion (\ref{EOMu}), (\ref{EOMv}) and Eq.\
(\ref{sigma}), we find
\begin{align}
\hbar {\cal S}_{\mathrm{spin}}&= \frac {\hbar \kappa S} 2 \int_0^{T_0}
\mathbf{C} \cdot \left( u \frac {\pd \bs \sigma} {\pd u} + v \frac {\pd \bs
\sigma} {\pd v} \right) \left(1 + |z|^2 \right) dt 
\nonumber \\
&= \hbar \kappa S \int_0^{T_0} \mathbf{C} \cdot \left(
\begin{array}{c}
u \left(1 - |z|^2 \right) \\
v \left(1 - |z|^2 \right) \\
2 |z|^2
\end{array}
\right) \frac {dt} {1+|z|^2}.
\label{spin}
\end{align}
Summing up the orbital and spin contributions Eqs.\ (\ref{orb}) and
(\ref{spin}), we obtain the entire change in action as
\begin{equation}
\delta {\cal S} = \delta {\cal S}_{pq} + \hbar {\cal S}_{\mathrm{spin}} = 
\hbar S \int_0^{T_0} F(t) dt,
\label{deltaS}
\end{equation}
where
\begin{equation}
F(t) = \kappa \mathbf{C} \cdot \left(
\begin{array}{c}
-u \\ -v \\ 1
\end{array}
\right).
\end{equation}

\subsection{Stability determinant} \label{stdet}

The stability determinant is derived from the second variation of the
Hamiltonian $H^{(2)}$ about the periodic orbit \cite{Ple}. In the leading order
in $\hbar$, the orbital and spin degrees of freedom in $H^{(2)}$ are separated.
This means that the spin phase space provides an additional block to the
unperturbed monodromy matrix of the orbital phase space, which results in a
separate stability determinant due to spin. The linearized momentum and
coordinate for spin 
\begin{equation}
\left(
  \begin{array}{c}
  \xi \\ \nu
  \end{array}
\right)
= \frac {2 \sqrt{\hbar S}} {1 + |z|^2} 
\left(
  \begin{array}{c}
  \delta u \\ \delta v
  \end{array}
\right)
\end{equation}
satisfy the equations of motion
\begin{equation}
\left(
  \begin{array}{c}
  \dot \xi \\ \dot \nu
  \end{array}
\right)
=
\left(
  \begin{array}{c}
  - \frac {\pd H^{(2)}} {\pd \nu} \vspace{.1cm}\\
  \frac {\pd H^{(2)}} {\pd \xi}
  \end{array}
\right)
= F(t)
\left(
  \begin{array}{c}
  - \nu \\ \xi
  \end{array}
\right).
\label{linEOM}
\end{equation}
Solving these equations we find the spin block of the monodromy matrix to be
(Appendix \ref{mon}) 
\begin{equation}
M =
\left(
  \begin{array}{cc}
  \cos \varphi & - \sin \varphi \\
  \sin \varphi &   \cos \varphi
  \end{array}
\right),
\label{monmat}
\end{equation}
where the stability angle is
\begin{equation}
\varphi = \int_0^{T_0} F(t) dt.
\end{equation}
The proportionality between $\varphi$ and $\delta {\cal S}$ [Eq.\
(\ref{deltaS})] will be exploited in a moment but, first, we find the
stability determinant
\begin{equation}
\left| \det (M - I) \right|^{1/2} = 2 \left| \sin \frac \varphi 2 \right|,
\end{equation}
where $I$ is the $2 \times 2$ unit matrix.

\subsection{Trace formula}

As was explained at the end of Sec.\ \ref{ClasDyn}, for each unperturbed
periodic orbit there are two new periodic orbits with opposite spin
orientations $\bs \sigma (t)$. It is easy to deduce, then, that for these two
orbits both $\delta {\cal S}$ and $\varphi$ have the same magnitude but
opposite signs. Now we are ready to write the trace formula for the oscillatory
part of the density of states
\begin{equation}
\delta g (E) = \sum_{po} \sum_{\pm} \frac {{\cal A}_0} {2 \left| \sin \frac
\varphi 2 \right|} \cos \left[ \frac 1 \hbar ({\cal S}_0 \pm \delta {\cal S}) -
\frac \pi 2 (\mu_0 + \mu_{\pm}) \right],
\label{trf}
\end{equation}
where the first sum is over the unperturbed periodic orbits and the second sum
takes care of the contribution of the two spin orientations; ${\cal A}_0$ is
the prefactor for the unperturbed orbit, which depends on the stability
determinant and the primitive period; ${\cal S}_0$ and $\mu_0$ are the
unperturbed action and the Maslov index, respectively; $\mu_{\pm}$ are the
additional Maslov indices due to spin. The nature of spin requires an
additional Kochetov-Solari phase correction \cite{Koc} that results in the
shift $S \longmapsto S + \frac 1 2$ of the total spin parameter in $\delta
{\cal S}$ (Appendix \ref{Koch}). Setting $S = \frac 1 2$, we end up with
\begin{equation}
\delta {\cal S} \longmapsto  \delta \tilde {\cal S} = \hbar \varphi.
\end{equation}
With this relation and the additional Maslov index (Appendix \ref{Mas})
\begin{equation}
\mu_{\pm} = 1 + 2 \left[ \pm \frac \varphi {2\pi} \right]
\label{Mind}
\end{equation}
($[x]$ is the largest integer $\leq x$) the sum over the spin orientations
in Eq.\ (\ref{trf}) becomes
\begin{align}
&\sum_{\pm} \frac {{\cal A}_0} {2 \sin \frac \varphi 2} \cos \left[ \left(
\frac  {{\cal S}_0} \hbar - \frac \pi 2 \mu_0 \right) \pm \left( \frac {\delta
\tilde {\cal S}} \hbar - \frac \pi 2 \right) \right] 
\nonumber \\
&= 2 \cos \left( \frac \varphi 2 \right) {{\cal A}_0} \cos \left[ \frac {{\cal
S}_0} \hbar - \frac \pi 2 \mu_0 \right].
\label{sum}
\end{align}
This is our main result: each term in the periodic orbit sum is the
contribution of an unperturbed orbit ${{\cal A}_0} \cos \left[ \frac {{\cal
S}_0} \hbar - \frac \pi 2 \mu_0 \right]$ times the modulation factor
\begin{equation}
{\cal M} = 2 \cos \left( \frac \varphi 2 \right).
\label{mf}
\end{equation}
Note that no assumption was made on whether the unperturbed periodic orbits are
isolated or not.

\section{Comparison with another method} \label{BK}

Bolte and Keppeler \cite{Bol} derived the modulation factor in the
weak-coupling limit by a different method. Their results\footnote{We
reformulate the results of Ref.\  \cite{Bol} for the south gauge.} are
expressed in terms of a spin trajectory with the initial condition 
\begin{equation}
\bs \sigma (0) = (0,0,-1)^{\mathrm{T}}
\label{incond}
\end{equation}
that obeys Eq.\ (\ref{EOM3}). This trajectory, in general, is not periodic. As
in our approach, the influence of spin on the orbital motion is neglected. The
spin motion can be described by the polar angles $\left(\theta(t),
\phi(t)\right)$ with $\theta(0) = \pi$. The modulation factor is then 
\begin{equation}
{\cal M}_{BK} = 2 \cos \left( \frac {\Delta \theta} 2 \right) \cos \chi,
\end{equation}
where $\Delta \theta = \pi - \theta(T_0)$ and\footnote{Ref.\ \cite{Bol} defines
the phase $\eta = - \chi$.}
\begin{equation}
\chi = - \frac \kappa 2 \int_0^{T_0} \mathbf{C} \cdot \bs \sigma dt + \frac
1 2 \int_0^{T_0} \left[ 1 + \cos \theta (t) \right] \dot \phi (t) dt.
\label{chi}
\end{equation}

In order to show that our modulation factor Eq.\ (\ref{mf}) is equal to ${\cal
M}_{BK}$, let us express $\varphi$ in terms of the polar angles. From Eq.\ 
(\ref{sigma}) follows the coordinate transformation
\begin{align}
&u = \cot \frac \theta 2 \cos \phi,
\nonumber \\
&v = \cot \frac \theta 2 \sin \phi.
\end{align}
Since  $\varphi \propto \delta {\cal S}$, we can represent it as a sum of two
terms [cf.\ Eqs.\ (\ref{orb})-(\ref{deltaS})]
\begin{equation}
\frac \varphi 2 = - \frac \kappa 2 \int_0^{T_0} \mathbf{C} \cdot \bs \sigma dt
+ \frac 1 2 \int_0^{T_0} \left[ 1 + \cos \theta (t) \right] \dot \phi (t) dt.
\label{phi}
\end{equation} 
There is a striking similarity between the expressions for $\chi$ and $\frac
\varphi 2$. The only difference is that in Eq.\ (\ref{chi}) the integration
is, in general, along a non-periodic orbit with the initial condition Eq.\ 
(\ref{incond}), while in Eq.\ (\ref{phi}) the integration is along the periodic
orbit. Since the modulation factor should not depend on the choice of the $z$
direction, we can choose the $z$ axis to coincide with the spin vector $\bs
\sigma (0)$ for the periodic orbit at $t=0$, i.e., the $z$ axis will be the
rotation axis in Fig.\ \ref{Fig1}. Then one of the periodic orbits will satisfy
the initial condition Eq.\  (\ref{incond}), and thus both $\chi$ and $\frac
\varphi 2$ can be calculated along this orbit and are equal. Moreover, $\Delta
\theta = 0$ in this case. Therefore the modulation factors derived within the
two approaches coincide,
\begin{equation}
{\cal M}_{BK} = {\cal M}.
\end{equation} 

It was mentioned in Ref.\ \cite{Kep} that ${\cal M}_{BK} = 2 \cos \frac \alpha
2$, where $\alpha$ is the rotation angle defined in Sec.\ \ref{ClasDyn}. Then,
of course, we conclude that
\begin{equation}
\cos \frac \alpha 2 = \cos \frac \varphi 2.
\label{alphi}
\end{equation} 
To see that this is indeed the case, we can go back to Sec.\ \ref{stdet} where
we calculated the stability determinant. It follows from that calculation that
the neighborhood of the periodic orbit is rotated by an angle $\varphi$ during
the period (Appendix \ref{mon}). Therefore the entire Bloch sphere is rotated
by this angle. Clearly, the angle of rotation must be defined $\mathrm{mod}\;
4\pi$, i.e., it depends  on the parity of the number of full revolutions of the
Bloch sphere around the periodic orbit during the period. It would be desirable
to prove Eq.\  (\ref{alphi}) without referring to the small neighborhood of the
periodic orbit. 

The same property can be also shown if one treats the spin quantum
mechanically. The spin propagator for the choice of the $z$ axis along the
rotation axis (so that $\chi = \frac \varphi 2$) is \cite{Bol}
\begin{equation}
d (T_0) = \left(
  \begin{array}{cc}
  e^{- i \frac \varphi 2} & 0\\
  0 & e^{i \frac \varphi 2} 
  \end{array}
\right).
\end{equation}
Applying this operator to a spinor $(\psi_+, \psi_-)^{\mathrm T}$ at $t = 0$,
we get the spinor $\left(\psi_+ e^{- i \frac \varphi 2}, \psi_- e^{i \frac
\varphi 2 }\right)^{\mathrm T}$ at $t = T_0$, which corresponds to the initial 
spin vector rotated by the angle $\varphi$ about the $z$ axis, i.e., $\varphi =
\alpha$.

\section{Summary and conclusions}

We have studied the case of weak spin-orbit coupling in the semiclassical
approximation using the spin coherent state method. The limit is achieved
formally by setting $\hbar \rightarrow 0$. The trajectories in the orbital
subspace of the extended phase space then remain unchanged by the spin-orbit
interaction. For each periodic orbit in the orbital subspace there are two
periodic orbits in the full phase space with opposite spin orientations. The
semiclassical trace formula can be expressed as a sum over unperturbed periodic
orbits with a modulation factor. This agrees with the results of Bolte and
Keppeler. The form of the modulation factor does not depend on whether the
unperturbed system has isolated orbits or whether it contains families of
degenerate orbits due to continuous symmetries. 

We remark that in the semiclassical treatment of pure spin systems, a
renormalization procedure is necessary in order to correct the stationary phase
approximation in the path integral for finite spin $S$. Such a renormalization
is equivalent to the Kochetov-Solari phase correction that we employed here
without justification for a system with spin-orbit interaction. Although this
correction worked well in our case, it may be necessary to develop a general
renormalization scheme when the interaction is not weak.

\section*{Acknowledgments}

The author thanks M. Pletyukhov and M. Brack for numerous constructive
discussions leading to this letter. This work has been supported by the
Deutsche Forschungsgemeinschaft.

\appendix

\section{Time parameterization} \label{time}

For pedagogical reasons we do the calculation in Eq.\ (\ref{deltaSpq}) with
the time parameterization. In this case ${\cal S}_{pq} = \int_0^T p \dot q dt$,
where $T$ is the exact period. Then the correction is
\begin{eqnarray}
\delta {\cal S}_{pq}&=& \int_0^{T_0} \left[ \delta p \dot q_0 + p_0 
\dot{(\delta q)} \right] dt + p_0 (T_0) \dot q_0 (T_0) \delta T
\nonumber \\
&=& \int_0^{T_0} \left( \delta p \dot q_0 - \delta q \dot p_0 \right) dt + 
p_0 \delta q \bigg|_0^{T_0} + p_0 (T_0) \dot q_0 (T_0) \delta T.
\end{eqnarray}
Transforming the boundary term 
\begin{align}
p_0 \delta q \bigg|_0^{T_0}&= p_0 (T_0) \left[ q (T_0) - q_0 (T_0) - q(0) +
q_0 (0) \right] = p_0 (T_0) \left[q (T_0) - q(0) \right]
\nonumber \\
&= p_0 (T_0) \left[q (T_0) - q(T) \right] \simeq - p_0 (T_0) \dot q_0 (T_0)
\delta T,
\end{align}
we see that it cancels the period correction term.

\section{Monodromy matrix} \label{mon}

We derive the monodromy matrix Eq.\ (\ref{monmat}). In order to solve the
equations of motion (\ref{linEOM}) we define $\eta = \xi + i \nu$. Then $\dot
\eta = i \eta F(t)$, which solves to
\begin{equation}
\eta (t) = \eta (0) \exp \left[ i \int_0^t F(t') dt' \right].
\end{equation}
It follows then that 
\begin{align}
&\xi (T_0) = \xi (0) \cos \varphi - \nu (0) \sin \varphi,
\nonumber \\
&\nu (T_0) = \xi (0) \sin \varphi + \nu (0) \cos \varphi,
\label{xinu}
\end{align}
resulting in Eq.\ (\ref{monmat}).

Note that according to Eq.\ (\ref{sigma}), 
\begin{align}
&\xi = \sqrt{\hbar S} \left( \delta \sigma_x + \frac {\sigma_x \delta \sigma_z}
{1 - \sigma_z} \right),
\nonumber \\
&\nu  = \sqrt{\hbar S} \left( \delta \sigma_y + \frac {\sigma_y \delta
\sigma_z} {1 - \sigma_z} \right).
\end{align}
If we choose the $z$ axis in such a way that the periodic orbit starts and ends
in the south pole, i.e., $\bs \sigma (0) = \bs \sigma (T_0) = (0,0,-1)^{\mathrm
T}$, then at $t = 0$ and $t = T_0$ we have 
\begin{align}
&\xi = \sqrt{\hbar S} \delta \sigma_x,
\nonumber \\
&\nu  = \sqrt{\hbar S} \delta \sigma_y.
\end{align}
Comparing with Eqs.\ (\ref{xinu}) we conclude that the neighborhood of the
periodic orbit is rotated by the angle $\varphi$ after the period.

\section{Kochetov-Solari phase shift} \label{Koch}

The Kochetov-Solari phase shift \cite {Koc} is given by 
\begin{equation}
\varphi_{KS} = \frac 1 2 \int_0^{T_0} A(t) dt,
\end{equation}
where
\begin{equation}
A(t) = \frac 1 {2 \hbar} \left[ \frac \pd {\pd \bar z} \frac {\left(1 +
|z|^2\right)^2} {2S} \frac {\pd H} {\pd z} + \mathrm{c.c.} \right].
\end{equation}
The spin-dependent part of the Hamiltonian is [cf.\ Eq.\ (\ref{Ham})]
\begin{equation}
\hbar H_{\mathrm{so}} (z, \bar z) = \frac {\hbar \kappa S} {1 + |z|^2}
\mathbf{C} \cdot \left(
  \begin{array}{c}
  z + \bar z \\ i (z - \bar z) \\ |z|^2 - 1
  \end{array}
\right).
\end{equation}
We find 
\begin{equation}
\frac \pd {\pd \bar z} \frac {\left(1 + |z|^2\right)^2} {2S} \frac {\pd
H_{\mathrm{so}}} {\pd z} = \kappa \mathbf{C} \cdot \left(
  \begin{array}{c}
  - \bar z \\ i \bar z \\ 1
  \end{array}
\right),
\end{equation}
therefore the phase shift becomes
\begin{equation}
\varphi_{KS} = \frac 1 2 \varphi.
\label{phiKS}
\end{equation}
Comparing to Eq.\ (\ref{deltaS}) we see that it effectively shifts the spin $S$
by $\frac 1 2$. One should keep in mind that this phase correction was
originally derived for a pure spin system. It has not been proven to have the
same form for a system with spin-orbit interaction. In the special case of the
weak-coupling limit we have a reason to believe that the result Eq.\ 
(\ref{phiKS}) is correct, since we were able to reproduce the modulation factor
found with another method \cite{Bol} (see Sec.\ \ref{BK}).

\section{Maslov indices} \label{Mas}

The additional Maslov indices $\mu_{\pm}$ are determined by the linearized spin
motion about the periodic orbit. The second variation of the Hamiltonian reads
[cf.\ Eq.\ (\ref{linEOM})]
\begin{equation}
H^{(2)} (\xi,\nu) = \frac {F(t)} 2 \left( \xi^2 + \nu^2 \right).
\end{equation}
Following Sugita \cite{Sug} we define its normal form 
\begin{equation}
H_{\mathrm{norm}} = \frac {\varphi} {2 T_0} \left( \xi^2 + \nu^2 \right)
\end{equation}
that has a constant frequency and generates the same phase change $\varphi$ as
$H^{(2)}$ after the period $T_0$. Then the spin block of the monodromy matrix
can be classified as elliptic and its Maslov index is given by Eq.\
(\ref{Mind}). $\varphi$ is the stability angle of one of the two orbits with
opposite spin orientations. Therefore, without loss of generality, we can
assume that $\varphi > 0$. Then, explicitly,
\begin{equation}
\mu_{\pm} = \left\{
  \begin{array}{ll}
  \pm 1, \;\; \mathrm{if} \;\; \varphi \in (0, 2\pi) &\mathrm{mod} \; 4\pi \\
  \pm 3, \;\; \mathrm{if} \;\; \varphi \in (2\pi, 4\pi) &\mathrm{mod} \; 4\pi
  \end{array}
\right. .
\end{equation}
On the other hand,
\begin{equation}
\mathrm{sign} \left( \sin \frac \varphi 2 \right) = \left\{
  \begin{array}{rll}
   1,& \;\; \mathrm{if} \;\; \varphi \in (0, 2\pi) &\mathrm{mod} \; 4\pi \\
  -1,& \;\; \mathrm{if} \;\; \varphi \in (2\pi, 4\pi) &\mathrm{mod} \; 4\pi
  \end{array}
\right. .
\end{equation}
Clearly, one can take $\mu_{\pm} = \pm 1$ and at the same time remove the
absolute value sign from $\sin \frac \varphi 2$, as was done in Eq.\
(\ref{sum}).

\end{document}